\documentstyle{article}\begin{document}
\title{Linear metric and temperature fluctuations of a charged plasma in a primordial magnetic field }
\author{ Z. Haba\\
Institute of Theoretical Physics, University of Wroclaw,\\ 50-204
Wroclaw, Plac Maxa Borna 9, Poland\\
email:zhab@ift.uni.wroc.pl}\maketitle
\begin{abstract} We discuss tensor metric perturbations in a magnetic field
around the homogeneous J\"uttner equilibrium of massless particles
in an expanding universe. We solve the Liouville equation and
derive the  energy-momentum tensor up to linear terms in the
metric and in the magnetic field. The term linear in the magnetic
field is different from zero if  the total charge of the
primordial plasma is non-zero.
 We obtain an analytic formula for  temperature
fluctuations treating the tensor metric perturbations and the
magnetic field as independent random variables. Assuming  a cutoff
on large momenta of the magnetic spectral function we show that
the presence of the    magnetic field can discriminate only  low
multipoles in the multipole expansion of temperature fluctuations.
In such a case the term linear in the magnetic field may be more
important than the quadratic one (corresponding to the
fluctuations of the pure magnetic field).
\end{abstract}

\section{Introduction}The magnetic field is ubiquitous in the
universe. In particular, the CMB results from quantum thermal
fluctuations of the electromagnetic field. It is present in the
standard model. Fluctuations of the magnetic field may be expected
in any model of the early universe. The non-trivial question
concerns  the appearance of a macroscopic magnetic field. There
are various mechanisms which can be responsible for this
phenomenon \cite{magrev}\cite{rev}. There is no convincing
argument for any of them. Let us mention the one which assumes a
non-zero total charge of the primordial plasma\cite{charge0}; the
assumption relevant for this paper.

The fluctuations lead to a diffusion of particle motions
\cite{habajpa} and to a random rhs of the Einstein equations
resulting from the energy-momentum. The energy-momentum contains a
contribution  of the free electromagnetic field. This term has
been studied in \cite{japan}\cite{durrer}\cite{mack}\cite{b2}.
When the magnetic field is Gaussian then the noise coming from the
energy-momentum of the free electromagnetic field being quadratic
is non-Gaussian. Non-Gaussian effects have not been discovered in
CMB yet. This may be so because the magnetic field is weak and the
quadratic terms are small. We point out in this paper that a
particle interaction with the magnetic field leads to a
 contribution to the energy-momentum of the primordial plasma which is linear in the
 magnetic field. This happens if the total charge of the primordial plasma is non-zero.
In such a case the impact of the linear term may be stronger than
 the one coming from the energy-momentum quadratic in the magnetic field.
The strength of this term depends on the charge. There are strong
bounds limiting the charge of the universe
\cite{jcap}\cite{charge}. We calculate a variation of the metric
corresponding to the term depending on the magnetic field. We
obtain an analytic formula for the temperature fluctuations
resulting from primordial fluctuations of the metric and of the
magnetic field. The temperature fluctuations contain an
information on structure formation. The impact of the primordial
magnetic field on structure formation (and temperature
fluctuations) is usually ignored. There are however some arguments
( see, e.g. \cite{peac},p.575) indicating that the magnetic field
should be taken into account in the studies of structure
formation.

The plan of this paper is the following. In sec.2 we find a
perturbative solution of the Liouville-Vlasov equation describing
a stream of particles  in an inhomogeneous expanding metric and in
the magnetic field. We are interested in the ultrarelativistic
limit when all the particles are massless. As a zero order
solution we choose the J\"uttner distribution \cite{juttner} with
a time dependent temperature. In sec.3 we discuss Einstein
equations with the energy-momentum on the rhs which is determined
by the solution of the Liouville-Vlasov equation. A perturbative
solution of Einstein equations determines a variation of the
metric in the magnetic field. Fluctuations of the temperature are
calculated in sec.4. Temperature fluctuations are expanded in
Legendre polynomials (multipole expansion). We study a dependence
of the expansion coefficients on the spectral function of the
stochastic primordial magnetic field. In the Appendix we discuss
some technical aspects of the estimates on the spectral function
of fluctuations of the magnetic field.

\section{Liouville-Vlasov equation}
In this section we solve perturbatively the Liouville-Vlasov
equation describing a distribution of classical trajectories (see
\cite{ellis}\cite{andrea} for its application in general
relativity). We decompose
\begin{equation}g_{\mu\nu}=\overline{h}_{\mu\nu}+h_{\mu\nu},
\end{equation}where
$\overline{h}_{\mu\nu}$ describes homogenous metric in the
conformal time and \begin{equation}
ds^{2}=g_{\mu\nu}dx^{\mu}dx^{\nu}=a^{2}(dt^{2}-d{\bf x}^{2}
-\gamma_{ij}dx^{i}dx^{j}).
\end{equation}
In eq.(2) we assume that the tensor perturbations $\gamma_{ij}$
are transverse and traceless. We write the Liouville equation in
the form
\begin{equation}
\begin{array}{l}
(p^{\mu}\partial^{x}_{\mu}-\overline{\Gamma}^{k}_{\mu\nu}p^{\mu}p^{\nu}\partial_{k})\Omega_{e}=
\delta\Gamma^{k}_{\mu\nu}p^{\mu}p^{\nu}\partial_{k}\Omega_{e}+eF^{j\nu}g_{\nu\mu}p^{\mu}\partial_{j}\Omega_{e},
\end{array}\end{equation}where $e$ is the electric charge,
 $\Gamma^{\mu}_{\nu\rho}$ are Christoffel symbols, $\partial_{\mu}^{x}=\frac{\partial}{\partial x^{\mu}}$ and
$x=(t,{\bf x})$ (boldface letters denoting the three vectors),
$\partial_{k}=\frac{\partial}{\partial p^{k}}$ denotes derivatives
over momenta.

For massless particles ( $m=0$) and in the homogeneous metric
($h_{\mu\nu}=0$)  any function $f(a^{2}\vert{\bf p}\vert)$ is a
solution of eq.(3)($\delta\Gamma=0$)\cite{bernstein}. Because of
the thermodynamic interpretation we choose the equilibrium
distribution $\Omega_{E}$ \cite{juttner} as a starting point of
the perturbation
\begin{equation} \Omega^{\mu}_{E}=g(2\pi)^{-3}\Big(\exp(a^{2}\beta\vert{\bf
p}\vert-\beta\mu(e))+q\Big)^{-1}
\end{equation}with\begin{displaymath}
{\bf p}^{2}=\sum_{j}p^{j}p^{j}.
\end{displaymath}$g$ is the number of particle's degrees of freedom (we set $g=1$ from now on), $\beta\equiv\frac{1}{T}$
is the inverse temperature
 and $\mu(e)$ is the chemical potential  for
particles of the type $e$. In eq.(4) $q=1$ for fermions, $q=-1$
for bosons and $q=0$ for the classical J\"uttner distribution. The
physical momentum is $a\vert{\bf p}\vert$ and the physical
temperature $(\beta a)^{-1}$.

 We write
\begin{equation} \Omega_{e}=\Omega_{E}^{\mu}(1+\chi_{e})\end{equation} and
look for a perturbative solution $\chi$ of eq.(3) in an
inhomogeneous metric (2).
 Then,

\begin{equation}\begin{array}{l}
\partial_{t}\chi+n^{k}\partial_{k}^{x}\chi- 2{\cal H}p^{k}\partial_{k}\chi
+e\vert{\bf
p}\vert^{-1}F^{j\nu}h_{\nu\mu}p^{\mu}\partial_{j}\chi\cr= -a^{2}\beta
f \vert {\bf
p}\vert(\partial_{t}\gamma_{jk}n^{j}n^{k}+\frac{1}{2}n^{j}\partial_{j}^{x}\gamma_{lk}n^{l}n^{k})+ea^{2}\beta
f \vert{\bf p}\vert^{-2}F^{j\nu}h_{\nu\mu}p^{\mu}p^{j},
\end{array}\end{equation}where $n^{k}=p^{k}\vert{\bf
p}\vert^{-1}$,  ${\cal H}=a^{-1}\frac{da}{dt}$ and
$f=\hat{\Omega}^{\prime}(a^{2}\beta \vert {\bf p}\vert)$ where
\begin{displaymath}
\hat{\Omega}^{\prime}(x)=\exp(x-\mu(e)\beta)\Big(\exp(x-\mu(e)\beta)+q\Big)^{-1}.
\end{displaymath}
$f=1$ for the J\"uttner distribution (relativistic equilibrium
distribution neglecting the quantum statistics).

We set $f\simeq 1$ and  look for a solution of eq.(6) which is of
the first order in momentum
\begin{equation}
\chi_{e}=a^{2}\beta\nu\vert{\bf p}\vert +\beta r_{e}.\end{equation}
 We assume an infinite conductivity of the primordial plasma. Then, the electric field is zero \cite{e0}. Inserting (7) in
eq.(6) we obtain  equations  for $\nu$ and
$r$\begin{equation}\begin{array}{l}
\partial_{t}\nu+n^{k}\partial_{k}^{x}\nu
=-\partial_{t}\gamma_{jk}n^{j}n^{k}-\frac{1}{2}n^{j}\partial_{j}^{x}\gamma_{lk}n^{l}n^{k},
\end{array}\end{equation}
\begin{equation}\begin{array}{l}
\partial_{t}r_{e}+n^{k}\partial_{k}^{x}r_{e}
=e\gamma_{km}\epsilon^{jkl}n^{j}n^{m}B_{l}\equiv \sigma_{e},
\end{array}\end{equation}
where we wrote
\begin{displaymath}
F^{jk}=\epsilon^{jkl}\tilde{B}_{l}(t)
\end{displaymath}
with
\begin{displaymath}
\tilde{B}(t,{\bf x})=a^{-2}B({\bf x})
\end{displaymath}
This time-dependence of the magnetic  field follows from Maxwell equations
in an expanding universe \cite{japmag}.
 We introduce \begin{equation}
\Theta=\nu-\frac{1}{2}\gamma_{jk}n^{j}n^{k}
\end{equation}Then
\begin{equation}\begin{array}{l}
\partial_{t}\Theta+n^{k}\partial_{k}^{x}\Theta
=-\frac{1}{2}\partial_{t}\gamma_{jk}n^{j}n^{k}\equiv {\cal R}
\end{array}\end{equation}
$\Theta$ as discussed in our earlier paper \cite{habampa} has the
meaning of the temperature variation. The solution of eq.(11)
reads \begin{equation}\begin{array}{l} \Theta_{t}({\bf
x})=\int_{0}^{t}ds {\cal R}(s,{\bf x}-(t-s){\bf
n}),\end{array}\end{equation} The Fourier transform of the
solution of eq.(9) is
\begin{equation}\begin{array}{l}
\tilde{r}_{e}({\bf k},t)=\int_{0}^{t}ds\tilde{\sigma}_{e}({\bf k},{\bf
n},s)\exp(-i{\bf kn}(t-s))\end{array}\end{equation}
 We calculate (in the conformal time) the
energy-momentum tensor till the first order in the metric and the
magnetic  field perturbation ( we preserve the quadratic term
$T_{EM}$ of the free electromagnetic field for a later
comparison). For this purpose  we sum the densities of $+1$ and
$-1$ particles \cite{plasma}
\begin{equation}\begin{array}{l}
T^{jl}=(2\pi)^{-3}\sqrt{g}\int d{\bf
p}p_{0}^{-1}p^{j}p^{l}(\Omega_{+}+\Omega_{-}) +T_{EM}^{jl}\cr=\int
d{\bf
n}n^{j}n^{l}\Big((1+4\nu)(\Omega^{(+)}_{E}+\Omega^{(-)}_{E})+\beta
(r_{+}\Omega^{(+)}_{E}+r_{-}\Omega^{(-)}_{E})\Big)
+T_{EM}^{jl}\cr\equiv T_{0}^{jl}+\delta T^{jl}+ T_{EM}^{jl}
\end{array}
\end{equation}where
\begin{displaymath}T_{EM}^{jl}= \frac{1}{8\pi
a^{6}}(2B^{j}B^{l}-\delta^{jl}B^{k}B^{k})
\end{displaymath}
Here, $T_{0}$ is the energy-momentum of  the solution (4) of the
Liouville  equation on the homogeneous space-time, $\delta T$
denotes the terms linear in the metric $\gamma$ and in the
magnetic field $B$. In eq.(14) $\nu$ should still be expressed by
$\Theta$ from eq.(10) or determined from eq.(8).

 We can calculate the tensor (14) taking
multiple derivatives

\begin{equation}
\int d{\bf  m}\exp(-i{\bf k}{\bf
m}(t-s))m^{j_{1}}....m^{j_{r}}=i^{r} \frac{\partial}{\partial
q^{j_{1}}}.....\frac{\partial}{\partial
q^{j_{r}}}q^{-1}\sin(q),\end{equation} where after the calculation
of derivatives we should set
\begin{displaymath}
q=(t-s)k.
\end{displaymath}

\section{Einstein equations in a magnetic field}

We write the spatial part of Einstein equations for traceless
transverse metric perturbations in the form ($G^{i}_{j}$ is the
Einstein tensor, $G$ is the Newton constant)
\begin{equation}
\begin{array}{l}
\delta G^{i}_{j}=-2a^{-2}\Big(
\frac{1}{2}\partial_{t}^{2}\gamma_{ij} -\frac{1}{2}\triangle
\gamma_{ij}  +{\cal H}\partial_{t}\gamma_{ij} \Big) \cr=8\pi a^{2}
G P_{ij;kl}(\delta T^{kl}+(T_{EM})^{lk}),\end{array}\end{equation}
where \begin{displaymath} P_{ij;kl}=\triangle
_{ik}\triangle_{lj}+\triangle _{il}\triangle_{kj}-\triangle
_{ij}\triangle_{lk},\end{displaymath}
\begin{displaymath}
\Delta_{jk}=\delta_{jk}-\partial_{j}\partial_{k}\triangle^{-1}
\end{displaymath}and $\triangle$ is the three-dimensional
Laplacian.
 The matter energy-momentum $\delta T$ on the
rhs is linearly dependent on the magnetic field whereas the free
electromagnetic energy-momentum $T_{EM}$ is quadratic in the
magnetic field.

 We write
\begin{equation}
\gamma_{jk}=a^{-1}\tilde{\gamma}_{jk}.
\end{equation}Then, eq.(16) takes the form
\begin{equation}{\cal G}\tilde{\gamma}_{jk}=
\partial_{t}^{2}\tilde{\gamma}_{jk}-(\triangle+a^{-1}\partial_{t}^{2}a )\tilde{\gamma}_{jk}=8\pi a^{5} G(
 (\delta T^{TT})_{jk}+(T_{EM}^{TT})_{jk}),\end{equation} where the
rhs still depends on the metric. For a general $a(t)$ it is not
simple to solve eq.(18). Let ${\cal G}^{-1}(k;t,s)$ be the kernel of the inverse of ${\cal G}$.
 Then, we can solve eq.(18) by iteration (till the first order in
 $G$)
\begin{equation}
\tilde{\gamma}_{jk}=\tilde{\gamma}_{jk}^{grav}(t)+8\pi
G\int_{0}^{t}ds {\cal G}^{-1}(k;t,s)a^{5} \delta
T^{TT}_{jk}(s,a^{-1}\tilde{\gamma}^{grav}),\end{equation} where
$\gamma^{grav}$ is the solution of the homogeneous equation ( at
$G=0$).

In the radiation era
$\partial_{t}^{2}a=0$. Then,
\begin{equation}
{\cal G}^{-1}(k;t,s)=k^{-1}\sin(k(t-s)).
\end{equation}
In another limit, if $ka<< \partial^{2}_{t}a$ then the dependence
of ${\cal G}^{-1}$ on $k$ can be neglected.

Returning to eq.(18) we perform some integrals over ${\bf n}$
(using eq.(15)) and write it in the form
\begin{equation}
\begin{array}{l}
 \partial_{t}^{2}\gamma_{ij} -\triangle
\gamma_{ij}  +2{\cal H}\partial_{t}\gamma_{ij} =8\pi Ga^{4}
P_{ij;kl}T_{EM}^{kl} \cr+8\pi G(2\pi)^{-3} 96\pi a^{-2}\beta^{-4}
P_{ij;kl} (\frac{4}{15}\gamma_{kl}+t^{lk}),
\end{array}\end{equation}
where
\begin{equation}t^{jl}=\int d{\bf
n}n^{j}n^{l}\Big(\Theta+\frac{1}{4}\beta
(N_{+}r_{+}+N_{-}r_{-})\Big)\equiv\theta^{jl}+r^{jl}.\end{equation}
$\Theta$ is determined from eqs.(11)-(12) and $r$ is defined in
(13) ($\theta$ and $r$  do not depend on $a$). For the equilibrium
distribution $\Omega_{\mu}^{E}$ we have assumed the approximate
formula (justified for high energies)
\begin{equation}
\Omega^{\mu}_{E}=N_{\mu}\exp(-\beta a^{2}\vert {\bf p}\vert),
\end{equation}where from eq.(4) at $q\rightarrow 0$
\begin{displaymath}N_{\mu}=\exp(\mu(e)\beta)\end{displaymath}

 In order
to write down Einstein equations explicitly we  calculate the part of the plasma  energy-momentum (22)
which is linear in the magnetic field
\begin{equation}\begin{array}{l} r^{jk}({\bf k},t)=(N_{+}-N_{-})\int_{0}^{t}
ds d{\bf m}m^{j}m^{k}\sigma({\bf k},{\bf m},s)\exp(-i{\bf
km}(t-s))\cr=(N_{+}-N_{-})\int_{0}^{t} ds d{\bf m}d{\bf
q}m^{j}m^{k}\gamma_{li}({\bf q},s)\epsilon^{rlp}B_{p}({\bf k}-{\bf
q})\exp(-i{\bf km}(t-s))m^{i}m^{r}.
\end{array}\end{equation}
${\bf m} $   denotes the directional vector of propagation (which
we denoted by ${\bf n}$ in eq.(9)). So, the part of the metric
perturbation coming from the magnetic field is

\begin{equation}
\begin{array}{l}
 \partial_{t}^{2}\gamma_{ij} -\triangle
\gamma_{ij}  +2{\cal H}\partial_{t}\gamma_{ij} =8\pi G(2\pi)^{-3}
96\pi a^{-2}\beta^{-4} P_{ij;kl} r^{lk}.
\end{array}\end{equation}

 For the remaining part of eqs.(21)-(22) we have
 \begin{displaymath} \delta T^{jk}=(2\pi)^{-3}96\pi
a^{-6}\beta^{-4}t^{jk}
\end{displaymath}
This part of metric fluctuations is discussed in many text-books
\cite{mukhanov} \cite{rubakov}(we have calculated it for a
diffusive matter in \cite{habampa}).
\section{Temperature fluctuations}There will be temperature fluctuations caused by the density
fluctuations (scalar perturbations), gravitational waves (quantum
metric fluctuations) as well as fluctuations of the primordial
magnetic fields. The solution for the temperature fluctuations is
expressed by $\gamma$ (eq.(12))
\begin{displaymath}\begin{array}{l}\Theta(t,{\bf n})=-n^{l}n^{j}
\frac{1}{2}\int_{0}^{t}ds\partial_{s}\gamma_{jl} (s,{\bf
x}-(t-s){\bf n}),\end{array}\end{displaymath} where $\gamma$ is
determined by $t^{jl}$.  Now,
\begin{equation}
\begin{array}{l}\langle \Theta(t,{\bf n})\Theta(t,{\bf n}^{\prime})\rangle=
 \cr=\frac{1}{4}
(2\pi)^{-3}\int_{0}^{t}ds\int_{0}^{t}ds^{\prime}\int d{\bf k}d{\bf
k}^{\prime}n^{l}n^{j}n^{\prime r}n^{\prime p}
\cr\partial_{s}\partial_{s^{\prime}}\langle\gamma_{jl} (s,{\bf
k})\gamma_{rp} (s^{\prime},{\bf k}^{\prime})\rangle
\exp(-i(t-s){\bf nk}+i(t-s^{\prime}){\bf n}^{\prime}{\bf k}),
\end{array}\end{equation}In eqs.(25)-(26) we wish to calculate the part of fluctuations
coming from the magnetic field. It is determined from the solution
of Einstein equations (16). From the perturbative solution (19)
the fluctuations coming from the magnetic field are
\begin{displaymath}\begin{array}{l} \langle\gamma_{jl} (s,{\bf
k})\gamma_{ab} (s^{\prime},{\bf k}^{\prime})\rangle=\Big(8\pi
G(2\pi)^{-3} 96\pi a^{-2}\beta^{-4}\Big)^{2}(a(s)a(s^{\prime}))^{-1}
\int_{0}^{s}\int_{0}^{s^{\prime}}d\tau d\tau^{\prime}\cr
a(\tau)a(\tau^{\prime}){\cal
G}^{-1}(k,s,\tau ){\cal
G}^{-1}(k^{\prime},s^{\prime},\tau^{\prime})P_{jl;pq}
P_{ab;mn}\langle r_{pq}(a^{-1}\tilde{\gamma},{\bf k},\tau)r_{mn}
(a^{-1}\tilde{\gamma},{\bf k}^{\prime},\tau^{\prime})\rangle
,\end{array}\end{displaymath} where $r$ is  expressed by the
metric and by the magnetic field in eq.(24).

 We have two random
fields in the solution $\Theta$: $\gamma$ and ${\bf B}$. We assume
that ${\bf B}$ is a  Gaussian random field with the covariance
\begin{equation}
\langle B_{i}({\bf k})B_{j}({\bf
k}^{\prime})\rangle=\Delta_{ij}({\bf k})\delta({\bf k}+{\bf
k}^{\prime})P_{B}({\bf k}),
\end{equation}where
\begin{displaymath}
\triangle_{jl}({\bf k})=\delta_{jl}-k_{j}k_{l}{\bf
k}^{-2}.\end{displaymath} $B({\bf k})$ is time-independent as
explained below eq.(9) ( we denote a function and its Fourier
transform by the same letter; the meaning should follow from the
context),
 $\gamma$ is  an independent random field with
the covariance
\begin{displaymath}\begin{array}{l}
\langle\gamma_{jl} (s,{\bf k})\gamma_{ab} (s^{\prime},{\bf
k}^{\prime})\rangle=P_{jl;ab}\delta({\bf k}+{\bf
k}^{\prime})P_{\gamma}(k;s,s^{\prime}).
\end{array}\end{displaymath}

 We wish to calculate the correction to the
temperature fluctuations coming from the interaction with the
magnetic field. Let us define
\begin{equation}
P_{\sigma}({\bf q},s,s^{\prime})\delta({\bf q}+{\bf
q}^{\prime})=\langle \sigma(s,{\bf q})\sigma(s^{\prime},{\bf
q}^{\prime})\rangle,
\end{equation}where

\begin{equation}\begin{array}{l}
\sigma({\bf k},{\bf m},s)=\int d{\bf p}\gamma_{lk}({\bf
p},s)\epsilon^{rla}B_{a}({\bf k}-{\bf p})m^{k}m^{r}
\end{array}\end{equation}  Then,\begin{equation}\begin{array}{l}
\langle\sigma({\bf k},{\bf m},s)\sigma({\bf k}^{\prime},{\bf
m}^{\prime},s^{\prime})\rangle=\int d{\bf p}d{\bf
p}^{\prime}\langle\gamma_{li}({\bf p},s)\epsilon^{rla}B_{a}({\bf
k}-{\bf p})m^{i}m^{r}\cr \gamma_{l^{\prime}i^{\prime}}({\bf
p^{\prime}},s^{\prime})\epsilon^{r^{\prime}l^{\prime}a^{\prime}}B_{a^{\prime}}({\bf
k}^{\prime}-{\bf p}^{\prime})m^{\prime r^{\prime}}m^{\prime
i^{\prime}}\rangle\cr =\delta({\bf k}+{\bf
k}^{\prime})\Delta_{aa^{\prime}}({\bf
k})P_{li;l^{\prime}i^{\prime}}({\bf k})
\epsilon^{rla}m^{i}m^{r}\epsilon^{a^{\prime}r^{\prime}l^{\prime}}m^{\prime
r^{\prime}}m^{\prime i^{\prime}} \int d{\bf p}P_{\gamma}({\bf
p};s,s^{\prime})P_{B}({\bf k}-{\bf p})\cr\equiv\delta({\bf k}+{\bf
k}^{\prime})w({\bf k},{\bf m},{\bf m}^{\prime})\int d{\bf
p}P_{\gamma}({\bf p};s,s^{\prime})P_{B}({\bf k}-{\bf p}),
\end{array}\end{equation}
where
\begin{displaymath} w({\bf k},{\bf m},{\bf
m}^{\prime})=\Delta_{aa^{\prime}}({\bf
k})P_{li;l^{\prime}i^{\prime}}({\bf k})
\epsilon^{rla}m^{i}m^{r}\epsilon^{a^{\prime}r^{\prime}l^{\prime}}m^{\prime
r^{\prime}}m^{\prime i^{\prime}}.\end{displaymath}  The spectral
function defining the fluctuations of eq.(26) is determined by
\begin{equation} P_{\sigma}({\bf k};s,s^{\prime})=\int d{\bf p}P_{\gamma}({\bf
p};s,s^{\prime})P_{B}({\bf k}-{\bf p}).\end{equation} We consider
a simplified version of the graviton correlation function
\begin{equation}
P_{\gamma}(k;s,s^{\prime})=f(s,s^{\prime})k^{-\alpha}
\end{equation} with $0\leq\alpha\leq 3$ and \begin{equation}
P_{B}(k)=b^{2}k^{\sigma}\exp(-\frac{k^{2}}{\lambda})
\end{equation} (usually with $-3<\sigma\leq 2$)
as a model for the spectral function of a primordial magnetic
field with a Debye frequency cutoff for $k^{2}>\lambda$
\cite{japmag}\cite{jedam}. We discuss the spectral function
$P_{\sigma}$ in the Appendix. We show that for a small $k$ the
spectral function $P_{\sigma}(k;s,s^{\prime})$ tends to $K
f(s,s^{\prime})$ (with a certain constant $K$) whereas for a large
$k$ it behaves like the graviton spectral function,i.e.,as
$k^{-\alpha}$. Hence, the magnetic field substantially changes the
powerlike behaviour of the power spectrum for a small $k$ but it
does not change the leading behaviour for a large $k$. For
comparison  the contribution of the pure electromagnetic
energy-momentum tensor to the spectral function of the temperature
fluctuations  is determined by

\begin{equation}
P_{2B}({\bf k})=\int d{\bf p}P_{ B}({\bf p})P_{B}({\bf k}-{\bf
p})\end{equation} We show in the Appendix that for $\sigma \geq
-1$ it tends to a constant  for a small $ k$ and decays
exponentially for $ k^{2}>\lambda$. So, it does not contribute to
high multipoles. For low multipoles its contribution behaves  as
$b^{4}$ (the fourth power of the strength of the magnetic field)
whereas the contribution to the temperature fluctuations (24) of
the term linear in the magnetic field is proportional to
$b^{2}(N_{+}-N_{-})^{2}$.
 Now, using (24),(26) and (30)-(31)
\begin{equation}\begin{array}{l}\langle\partial_{t}\gamma_{jl}({\bf k},t)\partial_{t^{\prime}}\gamma_{j^{\prime}l^{\prime}}({\bf
k}^{\prime},t^{\prime})\rangle=\Big(8\pi G(2\pi)^{-3} 96\pi
a^{-2}\beta^{-4}\Big)^{2} \delta({\bf k}+{\bf
k}^{\prime})\partial_{t}\partial_{t^{\prime}}\int_{0}^{t} ds
\int_{0}^{t^{\prime}} ds^{\prime} \cr
(a(t)a(t^{\prime}))^{-1}{\cal G}^{-1}(k;t,s){\cal
G}^{-1}(k^{\prime};t^{\prime},s^{\prime})a(s)a(s^{\prime})
\int_{0}^{s}  d\tau \int_{0}^{s^{\prime}} d\tau^{\prime}\cr
W_{jl;j^{\prime}l^{\prime}}(s-\tau,\tau,s^{\prime}-\tau^{\prime},\tau^{\prime};{\bf
k})P_{\sigma}(k;\tau,\tau^{\prime})\equiv \delta({\bf k}+{\bf
k}^{\prime})P_{jl;j^{\prime}l^{\prime}}({\bf
k})F(t,t^{\prime};{\bf k}),
\end{array}\end{equation}where
\begin{equation}\begin{array}{l}W_{jl;j^{\prime}l^{\prime}}(s-\tau,\tau,s^{\prime}-\tau^{\prime},\tau^{\prime};{\bf
k})=\int d{\bf m}d{\bf m}^{\prime} P_{jl;ab}({\bf
k})P_{j^{\prime}l^{\prime};a^{\prime}b^{\prime}}({\bf
k})m^{a}m^{b}m^{\prime a^{\prime}}m^{\prime b^{\prime}}
 \cr \exp(i{\bf k}{\bf m}(s-\tau)-i{\bf k}{\bf
m}^{\prime}(s^{\prime}-\tau^{\prime}))w({\bf k},{\bf m},{\bf
m}^{\prime})=\cr \int d{\bf m}d{\bf m}^{\prime} P_{jl;ab}({\bf
k})P_{j^{\prime}l^{\prime};a^{\prime}b^{\prime}}({\bf
k})m^{a}m^{b}m^{\prime a^{\prime}}m^{\prime b^{\prime}}
 \exp(i{\bf k}{\bf m}(s-\tau)-i{\bf k}{\bf
m}^{\prime}(s^{\prime}-\tau^{\prime}))\cr
\Delta_{aa^{\prime}}({\bf k})P_{li;l^{\prime}i^{\prime}}({\bf k})
\epsilon^{rla}m^{i}m^{r}\epsilon^{a^{\prime}r^{\prime}l^{\prime}}m^{\prime
r^{\prime}}m^{\prime i^{\prime}}.
\end{array}\end{equation} The ${\bf m}$ integrals can be evaluated from the formula

\begin{displaymath}\begin{array}{l}\int d{\bf m}
\exp(i{\bf k}{\bf m}(s-\tau))m^{a}m^{b}m^{i}m^{r}
\cr=(s-\tau)^{-4}\frac{\partial}{\partial
k_{a}}\frac{\partial}{\partial k_{b}}\frac{\partial}{\partial
k_{i}}\frac{\partial}{\partial
k_{r}}(k(s-\tau))^{-1}\sin(k(s-\tau)).
\end{array}\end{displaymath}

We have
\begin{displaymath}
P_{jl;ab}P_{jl;ab}=6.
\end{displaymath}
Hence, in
eq.(35)\begin{equation}\begin{array}{l}F(t,t^{\prime};{\bf
k})=\frac{1}{6}P_{jl;j^{\prime}l^{\prime}}
\partial_{t}\partial_{t^{\prime}}\int_{0}^{t} ds
\int_{0}^{t^{\prime}} ds^{\prime} (a(t)a(t^{\prime}))^{-1}{\cal
G}^{-1}(k;t,s){\cal G}^{-1}(k;t^{\prime},s^{\prime})
a(s)a(s^{\prime})\cr \int_{0}^{s} d\tau \int_{0}^{s^{\prime}}
d\tau^{\prime}
W_{jl;j^{\prime}l^{\prime}}(s-\tau,\tau,s^{\prime}-\tau^{\prime},\tau^{\prime};{\bf
k})P_{\sigma}(k;\tau,\tau^{\prime}).
\end{array}\end{equation}
 $F(t,t;k)$ is the spectral function for
the magnetic contribution to the temperature fluctuations. For  a
small $k$ it tends to $ g(t)$ (with a certain function $g$)
because $P_{\sigma}(k)\rightarrow K f(s,s^{\prime})$, and the
functions $W_{jl;j^{\prime}l^{\prime}}$ and ${\cal G}^{-1}(k;t,s)$
in eq.(37) for a small $k$ also tend to a constant multiplied by a
function of time. For a large $k$ the  spectrum distribution
$P_{\sigma}$
 behaves as $P_{\gamma}$ (the one for gravitons).  In the higher orders of the conventional perturbative calculations of
the temperature fluctuations
 for a large $k$ we would obtain (from eq.(19), with the
energy-momentum $T$ for matter fields on the rhs)
 the  contribution to temperature
 fluctuations similar to the one resulting from eq.(37). Hence, we can
 conclude that the magnetic field does not substantially modify the behaviour of
 the spectral function for large $k$ in comparison to the one
 without the magnetic field. Hence, it would not be detectable by a measurement of
large multipoles.

We do not have an explicit formula for ${\cal G}^{-1}$. However,
for a small $k$ , such that $k<< a^{-1}\partial_{t}^{2}a$, the
dependence on $k$ can be neglected. In such a case, eq. (37) gives
an analytic formula for metric fluctuations caused by a linear
dependence on the primordial magnetic field. After the analytic
calculation of the spectral function $F(t,t;k)$ in eq.(37) we are
able to derive a formula for the contribution of the primordial
magnetic field to the temperature fluctuations (26). We will be
brief in the discussion of this derivation because it is already
standard and described in many textbooks
\cite{mukhanov}\cite{rubakov}. We follow the calculations of our
earlier paper \cite{habampa} (concerning dissipative systems). In
the integral $d{\bf k}=dkk^{2}d{\bf e}$ we integrate first over
${\bf e}$ in the exponential in eq.(26). We obtain
\begin{equation} \begin{array}{l}\int d{\bf e}\exp(-i(t-s){\bf nk}+i(t-s^{\prime}){\bf n}^{\prime}{\bf k})\cr=2\pi
k^{-1}\vert (t-s){\bf n}-(t-s^{\prime}) {\bf n}^{\prime}\vert^{-1}
 \sin\Big(\vert(t-s)k{\bf
n}-(t-s^{\prime})k{\bf n}^{\prime}\vert\Big).
\end{array}\end{equation}Next, we use the expansion
\begin{equation}\begin{array}{l}
k^{-1}\vert(t-s){\bf n}-(t-s^{\prime}){\bf n}^{\prime}\vert^{-1}
 \sin\Big(\vert(t-s)k{\bf
n}-(t-s^{\prime})q{\bf n}^{\prime}\vert\Big) \cr
=\sum_{l=0}^{\infty}(2l+1)j_{l}(k(t-s))j_{l}(k(t-s^{\prime}))P_{l}({\bf
n}{\bf n}^{\prime}).
\end{array}\end{equation}  $j_{l}$ is the Bessel spherical
function related to the Bessel function $J$ \cite{grad}
\begin{equation}
j_{l}(z)=\sqrt{\frac{\pi}{2z}}J_{l+\frac{1}{2}}(z)
\end{equation}
and $P_{l}$ are the Legendre polynomials.

 If  $F$ (37) is known
then owing to eqs.(26) and (34)  there remains to perform the
integrals over $s$ and $k$ \begin{equation}
\begin{array}{l}\langle \Theta(t,{\bf n})\Theta(t,{\bf n}^{\prime})\rangle=
 \cr=\frac{1}{4}
(2\pi)^{-3}\sum_{l=0}^{\infty}\int_{0}^{t}ds\int_{0}^{t}ds^{\prime}\int
d{\bf k}F(s,s^{\prime},k)(2({\bf n}\Delta({\bf k}){\bf
n}^{\prime})^{2}-({\bf n}\Delta({\bf k}){\bf n})({\bf
n}^{\prime}\Delta({\bf k}){\bf n}^{\prime}))\cr
j_{l}(k(t-s))j_{l}(k(t-s^{\prime}))(2l+1)P_{l}({\bf n}{\bf
n}^{\prime}),
\end{array}\end{equation}
where\begin{equation} {\bf n}\Delta({\bf k}){\bf n}^{\prime}={\bf
n}{\bf n}^{\prime}-{\bf k}^{-2}({\bf kn})({\bf
kn}^{\prime})\equiv\Delta({\bf nn}^{\prime},{\bf en},{\bf
en}^{\prime}),
\end{equation}
\begin{equation}
{\bf n}\Delta({\bf k}){\bf n}=1-{\bf k}^{-2}({\bf kn})^{2}\equiv
\delta({\bf en}).
\end{equation}
This formula is the starting point of calculations in
\cite{mukhanov} (see also our calculations in \cite{habampa}). The
expansion in Legendre polynomials reads
\begin{equation}
\begin{array}{l}\langle \Theta(t,{\bf n})\Theta(t,{\bf n}^{\prime})\rangle=
 \sum_{l=0}^{\infty}(2l+1)\tilde{D}_{l}(t,{\bf nn}^{\prime})P_{l}({\bf
n}{\bf n}^{\prime})=\sum_{l=0}^{\infty}(2l+1)C_{l}(t)P_{l}({\bf
n}{\bf n}^{\prime}),
\end{array}\end{equation}(where $\tilde{D}_{l}$ is the term in
front of $ (2l+1)P_{l}$ in eq.(41)). In eq.(44) $\tilde{D}_{l}
P_{l}$ still must be expanded in Legendre polynomials if the
coefficients $C_{l}$ are to be independent of the angle. We have
from eqs.(26),(36) and (41)\begin{equation}
\begin{array}{l}\tilde{D}_{l}=
\frac{1}{16\pi^{2}}\int_{0}^{t}ds\int_{0}^{t}ds^{\prime}\int_{0}^{\infty}
dk k^{2}F(s,s^{\prime},k)\cr\Big( 2\Delta({\bf
nn}^{\prime},-i\partial_{s},i\partial_{s^{\prime}})^{2}-\delta(-i\partial_{s}
)\delta(i\partial_{s^{\prime}})\Big)
j_{l}(k(t-s))j_{l}(k(t-s^{\prime})).
\end{array}\end{equation} Let us
consider only the  term without derivatives  in eq.(45) (denoted
$D_{l}$) resulting from the expansion\begin{displaymath}
2\Delta({\bf
nn}^{\prime},-i\partial_{s},i\partial_{s^{\prime}})^{2}-\delta(-i\partial_{s}
)\delta(i\partial_{s^{\prime}})=2({\bf n
n}^{\prime})^{2}-1+O(\partial_{s},\partial_{s^{\prime}}),
\end{displaymath} where $O$ is a polynomial of at least first
order in derivatives. The  terms in eq.(45) with derivatives  can
be calculated when  $D_{l}$ are known \cite{mukhanov}. We have
\begin{equation}\begin{array}{l}
D_{l}=\frac{1}{16\pi^{2}} (2({\bf n
n}^{\prime})^{2}-1)\int_{0}^{t}ds^{\prime}\int_{0}^{t}ds \cr
\int_{0}^{\infty}dqq^{2}F(s,s^{\prime},q)j_{l}(q(t-s))j_{l}(q(t-s^{\prime}))
\end{array}\end{equation}
 We make  an approximation for $F$ in eq.(37) which can be
 justified on the basis of the discussion following eq.(37)
\begin{equation}
F(s,s^{\prime},q)=g(s,s^{\prime})q^{-3+\epsilon},
\end{equation}with a certain function $g$,
where $\epsilon$ is different for a large $q$ and  for a small
$q$. The main contribution to the integrals (45) with the
spherical Bessel functions $j_{l}(rq)$ comes from $rq\simeq l$.
Hence, we can see that the behaviour of the integrals (45)
corresponding to the spectral function $F$ (47) at large $q$ is
responsible for the behaviour of $D_{l}$ at large $l$ and the
behaviour of $F$ at small $q$ corresponds to small $l$ in $D_{l}$.
 We can derive
an exact result for the integral (46) using the formula 6.574 of
\cite{grad}($\gamma<\sigma$)
\begin{equation}
\begin{array}{l}
\int_{0}^{\infty}dqq^{-2+\epsilon}J_{l+\frac{1}{2}}(\sigma
q)J_{l+\frac{1}{2}}(\gamma q)
=\frac{1}{2^{2-\epsilon}}\frac{\Gamma(l+\frac{\epsilon}{2})}{\Gamma(\frac{3}{2}-\frac{\epsilon}{2})\Gamma(l+\frac{3}{2})}
(\frac{\gamma}{\sigma})^{l}\sqrt{\gamma\sigma}\sigma^{\epsilon}\cr
F(l+\frac{\epsilon}{2},-\frac{1}{2}+\frac{\epsilon}{2},l+\frac{3}{2},\frac{\gamma^{2}}{\sigma^{2}}),
\end{array}\end{equation}where $F(\alpha,\beta,\gamma,z)$ denotes
the hypergeometric function. Applying eq.(48) we
obtain\begin{equation}\begin{array}{l} D_{l}(t)=\frac{1}{16\pi}
(2({\bf n
n}^{\prime})^{2}-1)\frac{1}{2^{2-\epsilon}}\frac{\Gamma(l+\frac{\epsilon}{2})}{\Gamma(\frac{3}{2}-\frac{\epsilon}{2})
\Gamma(l+\frac{3}{2})} \int_{0}^{t}ds\int_{0}^{s}ds^{\prime}
(\frac{t-s^{\prime}}{t-s})^{l} (t-s)^{\epsilon}\cr g(s,s^{\prime})
 F(l+\frac{\epsilon}{2},-\frac{1}{2}+\frac{\epsilon}{2},l+\frac{3}{2},\frac{(t-s^{\prime})^{2}}{(t-s)^{2}}).\end{array}\end{equation}
 The integral (46) can easily be calculated if  $F$ (47) is concentrated  at
 $s=s^{\prime}=s_{d}$. This case describes an instantaneous metric perturbation (the metric perturbation is
 limited to the moment $s_{d}$)
 corresponding to a sudden decoupling at $s=s_{d}$ from the last scattering surface \cite{mukhanov}\cite{rubakov}.
  In such a case
  $s=s^{\prime}=s_{d}$ in the argument of the hypergeometric
 function (49). We can obtain the
  value of the hypergeometric function at 1 using the formula
\begin{displaymath} F(\alpha,\beta,\gamma,1)=
\frac{\Gamma(\gamma)\Gamma(\gamma-\alpha-\beta)}{\Gamma(\gamma-\alpha)\Gamma(\gamma-\beta)}.
\end{displaymath}Then, if $\epsilon=3-\rho$
 \begin{equation} D_{l}\simeq
\Gamma(l+\frac{\epsilon}{2})(\Gamma(l+2-\frac{\epsilon}{2}))^{-1}\simeq
l^{1-\rho} \end{equation} Eq.(50)  shows that the magnetic field
is changing the behaviour of temperature  fluctuations at small
$l$. $D_{l}$ calculated with the magnetic field corrections do not
decrease with $l$ as they do for the inflationary
$P_{\gamma}\simeq k^{-3}$ (then $\epsilon=0$ or $\rho=3$). It is
known that there is a discrepancy between the theoretical and
observational multipole contributions at low $l$ to temperature
fluctuations. A part of it can come from the primordial magnetic
field. It follows that the part depending on the magnetic field
behaves as
\begin{equation}
\begin{array}{l}\langle \Theta(t,{\bf n})\Theta(t,{\bf n}^{\prime})\rangle_{B}\simeq
 \sum_{l=0}^{\infty}(2l+1)(2({\bf n
n}^{\prime})^{2}-1)\cr\Big(8\pi G(2\pi)^{-3} 96\pi
a(t_{d})^{-2}\beta^{-4}\Big)^{2}b^{2} (N_{+}-N_{-})^{2}d_{l}P_{l}({\bf
n}{\bf n}^{\prime}),
\end{array}\end{equation}with a certain slowly varying $d_{l}$.
In eq.(51) $t_{d}$ is the decoupling time, $d_{l}$ can be
calculated from eq.(37) and (41) by a numerical evaluation of the
integrals. It varies slowly with $l$. There is still the
contribution from the energy-momentum $T_{EM}$ of the magnetic
field (discussed in refs.\cite{japan}-\cite{b2}). This
contribution depends on $b^{4}G^{2}a^{-4}$ (the forth power of the
magnetic field strength). It is decreasing faster with $l$ because
the power spectrum $P_{2B}(k)$ is decreasing exponentially. The
ratio of the numerical contributions of the linear and the
quadratic ($T_{EM}$) terms depends on several parameters : the
total charge $N_{+}-N_{-}$, the strength of the magnetic field $b$
and the temperature $(\beta a(t_{d}))^{-1}$ at the decoupling.
There are no precise estimates of these parameters. However, from
the dependence of $d_{l}$ on $l$ we could infer the presence of
the primordial magnetic field and the charge of the universe.
\section{Summary} We
have derived an elementary formula for the particle's density
distribution resulting from the perturbative solution of the
Liouville-Vlasov equation. We have discussed a variation of the
distribution which is linear in the magnetic field. It seems that
this term has been ignored in the hitherto studies of the magnetic
field in the universe. The linear term is non-zero if the
primordial plasma is charged. There are strict estimates on the
charge of the universe \cite{jcap} \cite{charge}. The most
elementary bound results from the argument that the electric
repulsion cannot be much bigger than the gravitational attraction.
This arguments restrict the ratio of the charge of the universe (in electron units) to
the baryonic number  to be of the order $10^{-18}$. The strength
of the magnetic field is also restricted to be extremely small:
1.0 nG in the epoch of the photon  last scattering  \cite{magrev}.
As a consequence the linear term gives a small contribution to the
temperature fluctuation spectrum. We have calculated its
dependence on the magnetic field. There are some undetermined
parameters in the formula. However, the functional form of the
temperature fluctuations could discriminate between various models
of the primordial magnetic field. The usually discussed quadratic
perturbation of Einstein equations resulting from the
electromagnetic energy-momentum has a contribution to the
temperature fluctuations which depends in a different way on the
probability distribution of the magnetic field and it does not
depend on the temperature at the decoupling.

{\bf Acknowledgement}

 The research is supported by NCN grant DEC-2013/09/B/ST2/03455

\section{Appendix}\setcounter{equation}{0} We wish to calculate
\begin{equation}
P_{\sigma}({\bf k})=\int d{\bf q}P_{\gamma}({\bf k}-{\bf
q})P_{B}({\bf q})\end{equation} with \begin{equation}
P_{\gamma}(k)=k^{-\alpha}
\end{equation}as a typical model for a graviton distribution and \begin{equation}
P_{B}(k)=k^{\sigma}\exp(-\frac{k^{2}}{\lambda})
\end{equation}for the magnetic field probability distribution.
Then, eq.(1) gives
\begin{equation}
\begin{array}{l}
P_{\sigma}(k)=2\pi\int_{0}^{\infty} dq
\int_{-1}^{1}dxq^{\delta+2}(k^{2}+q^{2}-2qkx)^{-\frac{\alpha}{2}}\exp(-\frac{q^{2}}{\lambda})
\cr=2\pi
k^{-1}(2-\alpha)^{-1}\int_{0}^{\infty}dqq^{\delta+1}\Big((k+q)^{2-\alpha}
-\vert k-q\vert^{2-\alpha}\Big)\exp(-\frac{q^{2}}{\lambda})
\cr=2\pi
k^{-1}(2-\alpha)^{-1}\int_{0}^{k}dqq^{\delta+1}\Big((k+q)^{2-\alpha}
-( k-q)^{2-\alpha}\Big)\exp(-\frac{q^{2}}{\lambda})\cr +2\pi
k^{-1}(2-\alpha)^{-1}\int_{k}^{\infty}dqq^{\delta+1}\Big((k+q)^{2-\alpha}
-(
q-k)^{2-\alpha}\Big)\exp(-\frac{q^{2}}{\lambda})\cr
=I_{1}(k)+I_{2}(k).\end{array}\end{equation}
It is easy to see that \begin{equation}
 I_{1}(k)=k^{-\alpha}g(k)
 \end{equation}
 where $\lim_{k\rightarrow \infty}g(k)=const\neq 0$
 and $I_{2}$ decays exponentially for a large $k$. Moreover, for
a small $k$ we obtain that $P_{\sigma}(0)>0$ is finite if
$\delta-\alpha>-3$ .

Concerning the convolution (34) of the magnetic spectral function
it is useful to consider its Fourier transform $\tilde{P}_{B}({\bf
y})$
\begin{equation}
P_{B}({\bf k})=\int d{\bf y}\tilde{P}_{B}({\bf y})\exp(i{\bf ky}).
\end{equation}
Applying the Fourier transform for the distribution (3) with
$\sigma=0$ we obtain
\begin{equation}
P_{B}(k)=\exp(-\frac{k^{2}}{\lambda})=\int d{\bf
y}\exp(-\frac{\lambda}{4}{\bf y}^{2})\exp(i{\bf
ky})(\pi\lambda)^{-\frac{3}{2}}.
\end{equation}Hence, the spectral function  defined in eq.(34) is
\begin{equation}\begin{array}{l}
P_{2B}(k)=\int d{\bf p}P_{B}({\bf p})P_{B}({\bf k}-{\bf p})
=(\frac{\lambda}{2})^{-3}\int d{\bf y}\exp(-\frac{\lambda}{4}{\bf
y}^{2})\exp(i{\bf ky})\cr\int d{\bf y}^{\prime}\delta({\bf y}-{\bf
y}^{\prime})\exp(-\frac{\lambda}{4}{\bf y^{\prime}}^{2})\exp(i{\bf
ky^{\prime}})=8(2\pi\lambda)^{\frac{3}{2}}\exp(-\frac{k^{2}}{2\lambda}).\end{array}
\end{equation}
For general $\sigma \neq 0$ it is difficult to obtain explicit
formulas for $P_{2B}$.  In general, in eq.(6) if $P_{B}$ is
decaying faster than any polynomial then also $\tilde{P}_{B}$ is
decaying in the same way and the convolution (34) of  such
functions has this property. From eq.(6) we can also derive the
relation
\begin{equation}
P_{2B}({\bf k}=0)=\int d{\bf y}(\tilde{P}_{B}({\bf y}))^{2}.
\end{equation}
Hence, if $\tilde{P}_{B}({\bf y})$ is square integrable then
$P_{2B}({\bf k})$ is regular at $k=0$ and disappears fast at large
$k$. The behaviour for small ${\bf k}$ is not generic with respect
of convolutions and Fourier transforms. We do not have special
reasons to apply the spectral functions exactly of the form (3).
In the literature a sharp cut-off at $\sqrt{\lambda}$ is also
applied \cite{paci}. For the properties of the stochastic magnetic
fields only the behaviour of $P_{B}$ for a large $k$ and
 a small $k$ is essential. We suggest to replace the spectral
function (3) by
\begin{equation}
P_{B}(k)=\int_{1}^{\infty}ds
s^{-1-\frac{\sigma}{2}}\exp(-s\frac{k^{2}}{\lambda}).
\end{equation}
This function has the same behaviour as the one in eq.(3) for
large as well as small $k$. It has the virtue that its Fourier
transform can easily be calculated and we obtain a workable
representation of $P_{2B}$. It can be shown by an explicit
calculation that if $\sigma \geq -1$ then $P_{2B}(0)\neq 0$ is
finite . It decays exponentially for a large $k$. If $\sigma<-1$
then $P_{2B}(k) $ is singular when $k\rightarrow 0$ but its
singularity is less than $k^{-\vert\sigma\vert}$ ( e.g., as
discussed in \cite{paci}, when $\sigma=-2$ then we have
$P_{2B}\simeq k^{-1}$ for a small $k$).

\end{document}